\documentclass[conference]{IEEEtran}
\ifCLASSINFOpdf
\else
\fi

\usepackage{cite}
\usepackage{amsmath,amssymb,amsfonts}
\usepackage{algorithmic}
\usepackage{graphicx}
\usepackage{textcomp}
\usepackage{xcolor}
\usepackage{pdflscape}
\usepackage{esint}
\usepackage{subfig}
\usepackage[ruled,vlined]{algorithm2e}

\begin{document}
%
\title{Online SLA Decomposition: Enabling Real-Time Adaptation to Evolving Network Systems}


\author{
\IEEEauthorblockN{Cyril Shih-Huan Hsu}
\IEEEauthorblockA{Informatics Institute\\
University of Amsterdam\\
Amsterdam, The Netherlands\\
s.h.hsu@uva.nl}
\and
\IEEEauthorblockN{Chrysa Papagianni}
\IEEEauthorblockA{Informatics Institute\\
University of Amsterdam\\
Amsterdam, The Netherlands\\
c.papagianni@uva.nl}
\\
\IEEEauthorblockN{Danny De Vleeschauwer}
\IEEEauthorblockA{Nokia Bell Labs\\
Antwerp, Belgium\\
danny.de\_vleeschauwer@nokia-bell-labs.com}
\and
\IEEEauthorblockN{Paola Grosso}
\IEEEauthorblockA{Informatics Institute\\
University of Amsterdam\\
Amsterdam, The Netherlands\\
p.grosso@uva.nl}
}


%


\maketitle

\begin{abstract}
When a network slice spans multiple technology domains, it is crucial for each domain to uphold the End-to-End (E2E) Service Level Agreement (SLA) associated with the slice. Consequently, the E2E SLA must be properly decomposed into partial SLAs that are assigned to each domain involved.
In a network slice management system with a two-level architecture, comprising an E2E service orchestrator and local domain controllers, we consider that the orchestrator has access only to historical data regarding the responses of local controllers to previous requests, and this information is used to construct a risk model for each domain.
In this study, we extend our previous work by investigating the dynamic nature of real-world systems and introducing an online learning-decomposition framework to tackle the dynamicity. We propose a framework that continuously updates the risk models based on the most recent feedback. This approach leverages key components such as online gradient descent and FIFO memory buffers, which enhance the stability and robustness of the overall process. Our empirical study on an analytic model-based simulator demonstrates that the proposed framework outperforms the state-of-the-art static approach, delivering more accurate and resilient SLA decomposition under varying conditions and data limitations. Furthermore, we provide a comprehensive complexity analysis of the proposed solution.
\end{abstract}


%
\IEEEpeerreviewmaketitle

\section{Introduction}

The fifth generation (5G) of mobile communication technology introduced a versatile, multi-service network designed to support a wide range of vertical industries with a diverse set of service requirements. In 5G and beyond, network slicing plays a pivotal role by enabling the establishment and management of multiple End-to-End (E2E) logical networks. These slices are built on shared infrastructure and are specifically customized to meet the particular requirements of a given service, which is outlined in Service Level Agreements (SLAs).
SLAs function as contracts between service providers and tenants, defining the expected Quality of Service (QoS) through well-defined, measurable benchmarks known as Service-Level Objectives (SLOs). These objectives encompass various performance metrics such as data throughput, latency, reliability, and security, among others.
A single network slice may traverse multiple segments of the network, including (radio) access, transport, and core networks, and it may involve collaboration between different operators and infrastructure providers. To ensure that the service meets the agreed-upon SLOs across these domains, it is essential to adjust the service parameters accordingly.
As a result, the E2E SLA linked to a network slice must be broken down into specific SLOs for each domain, allowing for effective resource allocation within each segment.
The problem has been described in~\cite{ietf-teas-5g-network-slice-application-03, hcltech2023networkslicing, iovanna2022networkslicing}. In~\cite{ietf-teas-5g-network-slice-application-03}, authors suggested decomposing the end-to-end requirement to the transport network requirement is one of the key issues in network slice requirement mapping. Authors in~\cite{hcltech2023networkslicing} outlined how slices are managed through lifecycle automation, orchestration, and real-time monitoring to ensure SLA compliance across different domains. In~\cite{iovanna2022networkslicing}, authors elaborate that E2E QoS depends on SLA parameters and that transport resources must align with the heterogeneous QoS needs of different slices.
The authors in~\cite{su2019resource} highlighted the importance of E2E SLA decomposition for resource allocation, while~\cite{10011552} indicates that AI-assisted SLA decomposition is key to automating complex 6G business.
\begin{figure}[h]
 \centering
\includegraphics[width=1.03\columnwidth]{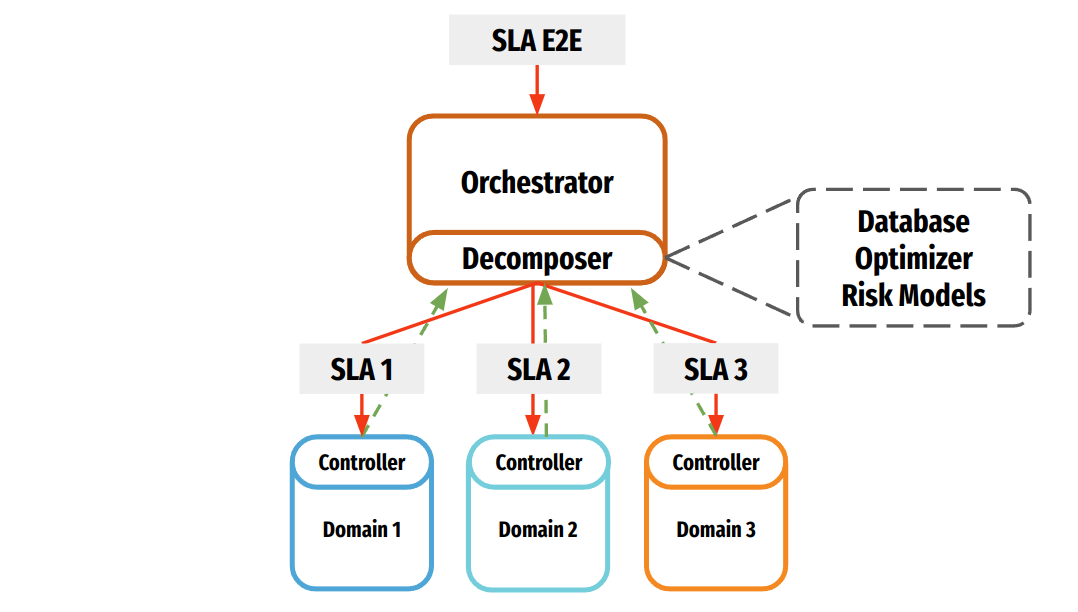}
\caption{Network slicing management and orchestration system, and the SLA decomposition module.}
\label{fig:NESMOS}
\vspace{-0.5em}
\end{figure}
Following the scenarios in~\cite{Vleeschauwer21_SLAdecomposition, SLADNN23}, in this paper we consider a two-level management architecture with an E2E service orchestrator handling network service lifecycle management, and local domain controllers managing slice instantiation within their domains, as shown in Fig.~\ref{fig:NESMOS}. The orchestrator determines the SLA decomposition for incoming service requests, while domain controllers handle admission control and resource allocation.
We assume the orchestrator lacks real-time knowledge of the infrastructure state but has access to historical feedback (i.e., request acceptance or rejection) from each domain. This allows the orchestrator to make informed decisions using domain-specific risk models based on the available data.


Some studies~\cite{8417711, 8931583, 10173672} have proposed prediction-based approaches for SLAs management, although they do not explicitly address E2E SLAs decomposition problem. In~\cite{8417711}, authors proposed the use of a mapping layer, which supervises the network over a service area and manages the allocation of radio resources to slices to guarantee their target service requirements.
Authors in~\cite{8931583} proposed a SLA-constrained optimization using Deep Learning (DL) to estimate the required resources based on the traffic per slice.
Another work~\cite{10173672} employs a context approach using graph representations for SLA violations prediction in Cloud Computing.
Moreover, several SLA decomposition methods employing heuristics have been studied~\cite{su2019resource}.
Authors in~\cite{9165317} present an E2E SLA decomposition system that applies supervised machine learning to break down E2E SLAs into access, transport, and core SLOs.
In our previous work~\cite{Vleeschauwer21_SLAdecomposition, SLADNN23}, we tackle the problem with a two-step approach, which is a combination of machine learning and optimization-based solution.

While the approaches in~\cite{Vleeschauwer21_SLAdecomposition, SLADNN23} have demonstrated success, they do not consider the inherent dynamicity of the system, which is critical for precise SLA management across network domains. This dynamicity is shaped by several factors, including variations in traffic intensity, shifts in user behavior, and fluctuating network conditions. As these factors evolve, the dynamics within each domain also change, potentially affecting the decision-making process of domain controllers. To address this gap, we propose an online learning-decomposition framework on top of~\cite{SLADNN23}, specifically tailored for SLA management in dynamic, multi-domain environments.

The contributions of this paper are summarized as:
\begin{enumerate}
\item [1.] We propose an online learning-decomposition framework that continuously updates risk models based on real-time feedback from domain controllers, enabling adaptation to dynamic conditions.
\item [2.] We leverage First In First Out (FIFO) memory buffers for data management to improved robustness and performance.
\item [3.] We conduct an ablation study to examine the contribution and importance of individual component of the proposed framework, and compare with the state-of-the-art offline learning approach, followed by a complexity analysis.
\end{enumerate}

The rest of this paper is organized as follows; Section \ref{sec:problem} presents background information and problem formulation on the SLA decomposition. In Section \ref{sec:methods} we present the proposed online learning-decomposition framework. Section \ref{sec:performance} details the experimental setup, followed by Section \ref{sec:results}, where we assess the performance and model complexity of the proposed framework. Finally, Section \ref{sec:conclusion} presents our conclusions.

\section{Problem Description}
\label{sec:problem}

\subsection{SLA Decomposition and Risk Models}

An E2E SLA, denoted by $s_{e2e}$, is a collection of SLOs linked to specific performance indicators. The SLO vector outlines the SLA's performance requirements in a sequential manner. For example, an SLA encompassing E2E delay and throughput is expressed as $s_{e2e}=(\tau_{e2e}, \theta_{e2e})$. This implies that the network slice must operate in a way that meets constraints imposed by $\tau_{e2e}$ for delay and $\theta_{e2e}$ for throughput, ensuring $\tau \leq \tau_{e2e}$, and $\theta \ge \theta_{e2e}$.
Considering a network slice distributed across $N$ domains (where n ranges from 1 to $N$), we introduce $s_n$ to represent the SLOs of the $n$-th domain. The relationship between individual domain SLOs and the overall E2E objective $s_{e2e}$ is defined by $s_{e2e}=G(s_1, s_2,..., s_N)$. For instance, the E2E delay is the sum of all delays for the involved domains, while the E2E throughput is determined by the lowest throughput across all domains. Mathematically, this is represented as:
\begin{equation}
    \begin{aligned}
    \tau_{e2e} &= \sum_{n=1}^{N}\tau_n,\\
    \theta_{e2e} &= \min\{\theta_1, \theta_2,..., \theta_N\}.
    \end{aligned}
\end{equation}

We can model the ability of a domain to support a partial SLA $s_n$ with a risk model, and the risk models of all involved domains can then be used in the SLA decomposition process.
The risk model is defined as $-\log P_n(s_n)$, where $P_n(s_n)$ represents the probability that a request in the $n$-th domain with SLOs $s_n$ is accepted. Under the assumption of independent decision-making by each domain, the overall E2E acceptance probability is calculated as the product of the individual acceptance probabilities of all involved domains. Therefore, the E2E decomposition can be formulated as an optimization problem that minimizes the overall risk in objective~(\ref{eq:obj}) under the constraints~(\ref{eq:composition}):
\begin{equation} \label{eq:obj}
-\sum_{n=1}^{N} \log P_n(s_n)
\end{equation}
\begin{equation} \label{eq:composition}
s_{e2e}=G(s_1, s_2,..., s_N)
\end{equation}


\subsection{Determining Neural Network-based Risk Models}
We determine the risk model per domain with a parameterized Neural Network (NN) $\mathcal{F}$~\cite{SLADNN23}, where the probability $P_n(s_n)$ is modelled as $\mathcal{F}_n(s_n)$.
When a domain is presented with a new service request with specific SLOs, denoted by $s$ (domain subscript omitted for simplicity), a controller must determine whether to accept or reject the request. This decision depends not only on the requested SLOs but also on the current infrastructure state, represented by $\omega$. The infrastructure state encompasses factors such as link and server utilization, network hop delays, and available backup paths, etc.
Therefore, the decision to accept or reject a service request depends jointly on the SLOs $s$ and infrastructure state $\omega$. While the domain controller has granular visibility of the infrastructure state, the orchestrator lacks this level of detail. Accordingly, although the controller's decision-making process is deterministic, the orchestrator perceives it as stochastic due to the unknown infrastructure state $\omega$.

Nevertheless, the acceptance probability $P$ can be estimated by analyzing the domain controller's responses to past requests. Given a set of $K$ responses to previous requests $\{(x_1, y_1), (x_2, y_2),..., (x_K, y_K)\}$, each represented by a proposed SLO $x$ and its corresponding acceptance decision $y$ (0 for rejection, 1 for acceptance), we can model the acceptance probabilities for SLO vectors with parameterized neural networks $\mathcal{F}$ by maximizing the overall likelihood on the dataset:
\begin{equation} \label{eq:risk}
\sum_{i=1}^{K}[y_i*\log(\mathcal{F}(x_i)) + (1-y_i)*\log(1-\mathcal{F}(x_i))].
\end{equation}
Furthermore, the acceptance probability exhibits a partial ordering relation~\cite{Vleeschauwer21_SLAdecomposition}, which incorporates the concept of SLA strictness, i.e., given a set of $K$ SLOs $S=\{x_1, x_2,..., x_K\}$, the acceptance probability has the following property:
\begin{equation} \label{eq:partial}
\forall x_i, x_j \in S,\, P(x_i) \leq P(x_j) \quad \textrm{if} \quad x_i \preceq x_j,
\end{equation}
The property indicates that a stricter SLO $x_i$ is less likely to be accepted compared to $x_j$. The authors in~\cite{SLADNN23} proposed several effective methods to bake this property into NN-based risk models without incurring any architectural constraints.

\subsection{Dynamicity}
The dynamicity of the system is crucial for accurately managing SLAs in network domains. Multiple factors contribute to this dynamicity, including traffic intensity fluctuations, changes in user behavior, varying network conditions and security threats. As these factors change, the acceptance probability of requests within a domain varies, impacting overall network performance.
Particularly, the acceptance probability becomes time-dependent. At a discrete time step $t$, the acceptance probability (in a single domain) of the SLO request $s$ is denoted as $P_t(s)$, which is influenced by the state $\omega_t$ of the domain, as stated in Section~\ref{sec:problem}-B. Given that $\omega_t$ evolves over time, the acceptance probability $P_t(s)$ also varies, even for the same SLOs. This behavior necessitates continuous learning and adaptation of the risk models.
Dynamic systems require periodic updates to risk models based on the recent feedback to ensure these models remain relevant and accurate. In the next section, we introduce a novel framework that allows the system to adapt to varying conditions and maintain near-optimal performance.
\section{Methodology}
In this section, we present the proposed framework \textbf{R}eal-time \textbf{A}daptive \textbf{DE}composition (RADE).
\label{sec:methods}
\subsection{Background}
A two-step decomposition approach was proposed in~\cite{Vleeschauwer21_SLAdecomposition, SLADNN23}. A dedicated risk model is learned first for each domain, given the historical feedback (i.e., proposed SLAs and their corresponding acceptance/rejection decisions made by controllers) from domain controllers. Once the risk models are built, an optimization proceeds to search for the decomposition that maximizes the E2E acceptance probability, as formulated in~(\ref{eq:obj}) and (\ref{eq:composition}).
The E2E SLA is decomposed into domain-specific SLAs that maximize the overall acceptance probability using a grid search followed by Sequential Least Squares Programming (SLSQP) algorithm.
However, in the previous work, the risk models are trained once and the weights are kept fixed for all future SLA decompositions. This approach is impractical for real-world systems, which are typically dynamic and time-dependent. The static models often lead to sub-optimal performance. We describe our extended solution to tackle this problem in the next subsection.
\begin{figure}[h]
 \centering
\includegraphics[width=0.95\columnwidth]{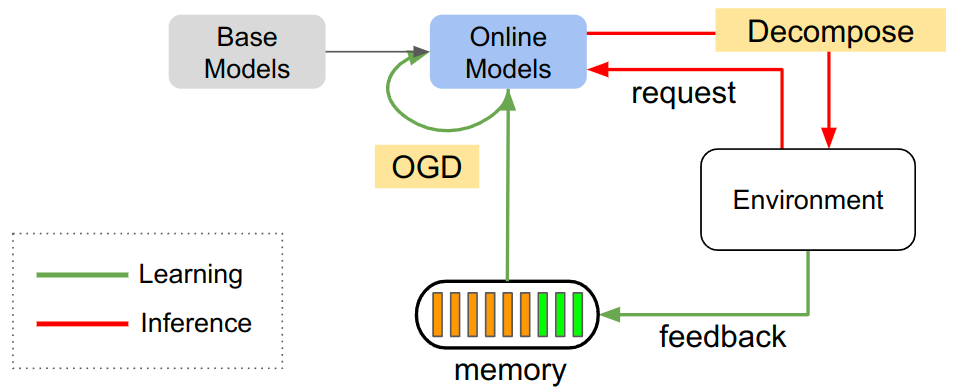}
\caption{Illustration of RADE framework.}
\label{fig:framework}
\end{figure} 
\subsection{Online Learning Framework}
To capture the dynamic nature of the environment, where the decision-making process of controllers evolves over time, it is essential to constantly update risk models based on the recent feedback. To this end, we propose an online learning-decomposition framework RADE, which is capable of running stable update as well as providing resilience against noisy samples. The following are the key components of RADE:
\noindent\textbf{Base model.} Following the design in~\cite{SLADNN23}, the base model is an NN. To account for monotonicity described in~(\ref{eq:partial}), we employ Absolute Weight Transformation (AWET) approach that shows prominent performance, namely $y = |w| \cdot x + b$ for $w, x, b, y$ being the weight, input, bias, and output of a neuron, respectively.
AWET ensures the weights remain non-negative, a sufficient condition for an NN to be monotonic, while still allowing the model's weights to be optimized freely.
\\
\noindent\textbf{Online update.} Unlike traditional static models, which are trained once and applied indefinitely, our approach involves periodic updates to the model based on the most recent feedback collected within each discrete time step. As illustrated in Fig.~\ref{fig:framework}, the loop begins with a base model and employs simple Online Gradient Descent (OGD)~\cite{HOI2021249} to perform updates. The continuously updated model is then used for real-time decomposition, ensuring that the system adapts promptly to the latest conditions.\\
\noindent\textbf{FIFO memory buffer.} Updating the model solely based on the most recent observations can lead to instability. For instance, the model may overfit when the feedback data is sparse, or learning may be compromised if feedback data contains errors. To mitigate these issues, we propose using a FIFO buffer with finite capacity for storing feedback. The FIFO buffer ensures a more stable and reliable learning process by maintaining a portion of historical feedback alongside all recent feedback~\cite{FIFO}. The limited capacity is necessary to ensure that outdated information is discarded, allowing the dataset to remain current and relevant for ongoing learning. This component helps prevent overfitting by providing a more diverse set of training samples and safeguards against the detrimental effects of occasional corrupted feedback.\\ 
\noindent\textbf{Online decomposition.} Besides the learning loop, the inference loop is also running in parallel. Upon receiving a new request, the associated E2E SLA will be decomposed with the latest risk models. The optimization-based decomposition follows the one proposed in~\cite{SLADNN23}. By leveraging up-to-date risk models, the inference loop ensures that the decomposition accurately reflects the current state of the corresponding domains.

\begin{algorithm}
\SetAlgoLined
\SetKwInput{kwParallel}{In parallel}
\KwIn{Step $\eta$, risk models $\mathcal{F}_{\theta}$, Memory buffer $\mathcal{M}$}
\For{$t = 1, 2, \ldots, T$}{

    \kwParallel {
        \Indp
        \\
        \Begin(Learning:){ 
        Receive a set of feedback $\mathcal{K}_t$\\
        $\mathcal{M} \gets \text{FIFO-PUSH}(\mathcal{M}, \mathcal{K}_t)$\\
        $\theta \gets \theta - \eta \nabla \mathcal{F}_{\theta}(\mathbf{\mathcal{M}})$ \tcp{multi-run} \\
        }
        \Begin(Inference:){ 
            Receive a set of requests $\mathcal{R}_t$\\
            with associated SLAs $\mathcal{S}_{e2e}$\\
            $\mathcal{S}_{partial} \gets \text{decompose}(\mathcal{F}_{\theta}, \mathcal{R}_t, \mathcal{S}_{e2e})$\\
            \Return $\mathcal{S}_{partial}$
        }
        \Indm
    }

}
\caption{RADE framework}
\label{alg:rade}
\end{algorithm}


Alg.~\ref{alg:rade} details the steps of the online update and decomposition mechanism.
The RADE framework operates in a time-stepped manner, where two key processes—learning and inference—run in parallel during each time step. At the beginning of each step, the framework receives feedback from the network regarding the performance of recent service requests. This feedback, represented as a set $\mathcal{K}_t$, is pushed into a FIFO memory buffer $\mathcal{M}$ to ensure that only the most recent and relevant data are stored. This buffer helps balance between old and new feedback. Using this stored feedback, the framework updates the risk models   through an online gradient descent process, adjusting the parameters based on the gradient of the loss function defined in~(\ref{eq:risk}). These continuously updated models help the system stay adaptive to changing network conditions. Simultaneously, the framework processes a set of new service requests $\mathcal{R}_t$, each accompanied by an E2E SLA $\mathcal{S}_{e2e}$ that defines the expected performance.
The E2E SLAs are then decomposed into partial SLAs $\mathcal{S}_{partial}$ for individual network domains by leveraging the up-to-date risk models $\mathcal{F}_{\theta}$ via optimizing:
\begin{equation}\label{eq:optimization_problem}
\begin{aligned}
\min_{s_n} \quad & -\sum_{n}^{N} \log f_{n, \theta}(s_n) \\
\text{s.t.} \quad & s_{e2e}=G(s_1, s_2,..., s_N), \\
\end{aligned}
\end{equation}
where $f_{n, \theta} \in \mathcal{F}_{\theta}$ represents the risk model corresponding to domain $n$, and $s_{e2e} \in \mathcal{S}_{e2e}$ denotes the E2E SLA associated with a certain request.
The decomposition process ensures that the E2E SLA are allocated across the network domains in an optimal manner such that the overall E2E acceptance probability is maximized.
In the subsequent sections, we will present the simulation environment for evaluations, followed by empirical results.
\section{Experimental Setup}
\label{sec:performance}

\subsection{Simulation Environment}
We follow the analytic model and data generation process proposed in~\cite{Vleeschauwer21_SLAdecomposition, SLADNN23} to generate data for three domains, which effectively cover the characteristics associated with URLLC, mMTC, and eMBB service requirements.
This analytic model maps a decomposition assignment to a probability, indicating how possible the given assignment will be accepted by the current domain controllers.
To introduce dynamicity into the system, we assume that the acceptance probability is inversely proportional to the current traffic intensity. Specifically, we define a time-dependent factor:
\begin{equation} \label{eq:lamb}
\lambda_t = \frac{1}{2}\left(\frac{\sin(2\pi t)}{T}+1\right) \cdot 0.9 + 0.1, 
\end{equation}
which represents the traffic intensity at time step $t$ over total number of steps $T$.
We then upgrade the form factor $\alpha$ proposed in~\cite{Vleeschauwer21_SLAdecomposition}, which models the probability distribution of the current load on the system, defined as:
\begin{equation}
    \alpha^\prime = \frac{\alpha}{\lambda_t},
\end{equation}
thereby accounting for the impact of varying load on the acceptance probability.
This adjustment reflects the changing traffic conditions over time. Note that the form factor $\alpha$ controls the level of acceptance probability: a larger $\alpha$ results in a higher acceptance probability for the same SLA request.
\subsection{Evaluation scenarios and metrics}
\noindent\textbf{Average acceptance probability over time.} We run Alg.~\ref{alg:rade} within the simulation environment detailed in Section~\ref{sec:performance}-A.
Following the assumption described in Section~\ref{sec:problem} that the decision-making processes of all domains are statistically independent, the E2E acceptance probability at each time step $t$ is calculated as the product of the individual acceptance probabilities across all three domains.
The average acceptance probability is then reported over the entire simulation period:
\begin{equation} \label{eq:avgprob}
p_{avg} = \frac{1}{T}\sum_{t=1}^T\frac{1}{M_t}\sum_{m=1}^{M_t}\prod_{d=1}^{D}P_{d, t}(s_{m,d,t}),
\end{equation}
where $T$ denotes the number of total time steps, $M_t$ is the number of requests at time $t$, $D$ is the number of involved domains (which is set to 3 in this paper), $P_{d, t}$ represents the analytic model, and $s_{m, d, t}$ is the decomposed partial SLA of the $m$-th request that is assigned to the $d$-th domain at time $t$. The number of requests $M_t$ at time $t$ is sampled from the Poisson distribution with $\lambda = \lambda_t$ described in~(\ref{eq:lamb}). The E2E SLA for each request is given as $(\tau_{e2e}, \theta_{e2e})$, where $\tau_{e2e}$ and $\theta_{e2e}$ are sampled uniformly from [90ms, 110ms] and [0.4Gbps, 0.6Gbps], respectively.\\
\noindent\textbf{Resilience test.} To assess the framework's resilience against corrupted feedback labels, we perform a resilience test on top of the aforementioned test, where each feedback has a corruption probability $p_c$ of being corrupted (i.e., the request is always rejected), which can result from issues like network delays, transient errors, or misconfiguration in domain controllers. Specifically, we track how the model's accuracy is affected as $p_c$ increases. This resilience test offers crucial insights into the effectiveness of the use of FIFO memory buffer mentioned in Section~\ref{sec:methods}-B, and the overall robustness of our framework under challenging conditions.

\subsection{Configurations}
To evaluate the contribution of each component proposed in Section~\ref{sec:methods}-B, we perform an ablation study by incrementally adding improvements at three stages, with each stage forming a distinct comparison method. Furthermore, two additional methods (Random and OPT) are included as benchmarks.

\noindent\textbf{Random.} The Random method does not employ any risk models and instead decomposes each incoming request's E2E SLA uniformly at random. This method is used as a baseline to verify the effectiveness of the proposed methods.\\
\noindent\textbf{Static.} The state-of-the-art method from our previous work~\cite{SLADNN23} involves a one-time training of risk models using feedback collected from a single prior run with the Random method. The weights of these risk models are then fixed and applied to all subsequent decompositions over time.
The model is a 3-layer Multi-Layer Perceptron (MLP), with 8 neurons each. The hyperbolic tangent (Tanh) activation function and Batch Normalization (BN) are applied for hidden layers in the order of linear-Tanh-BN.\\
\noindent\textbf{RADE/RADE*.} RADE represents the full method described in Section~\ref{sec:methods}-B, whereas RADE* is a variant of RADE that omits the FIFO memory buffer. In RADE*, risk models are updated using only the most recent feedback. Both RADE and RADE* utilize the Static method to initialize the risk models.\\
\noindent\textbf{OPT.} An exhaustive search is conducted at every time step to find the decomposition assignment that yields the largest E2E acceptance probability. This method provides the maximum theoretical performance achievable.

\section{Results and Discussion}
\label{sec:results}

\subsection{Performance}
\begin{figure}[h]
 \centering
\includegraphics[width=0.9\columnwidth, trim={0 0 0 0.5cm}, clip]{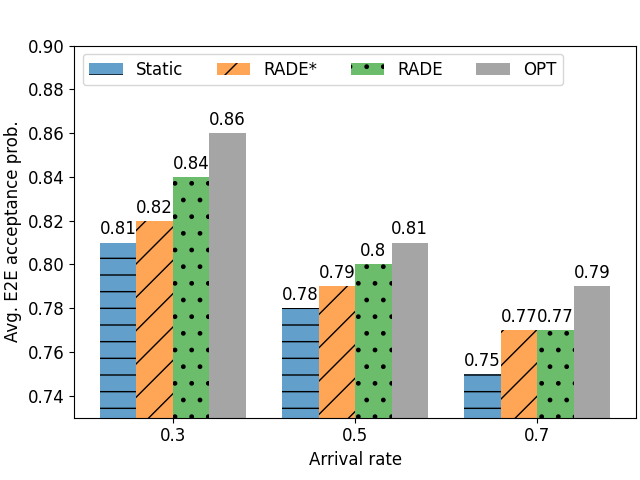}
\caption{Average E2E acceptance probability vs. arrival rate.}
\label{fig:am_main}
\end{figure}

\begin{figure}[b]
 \centering
\includegraphics[width=1\columnwidth]{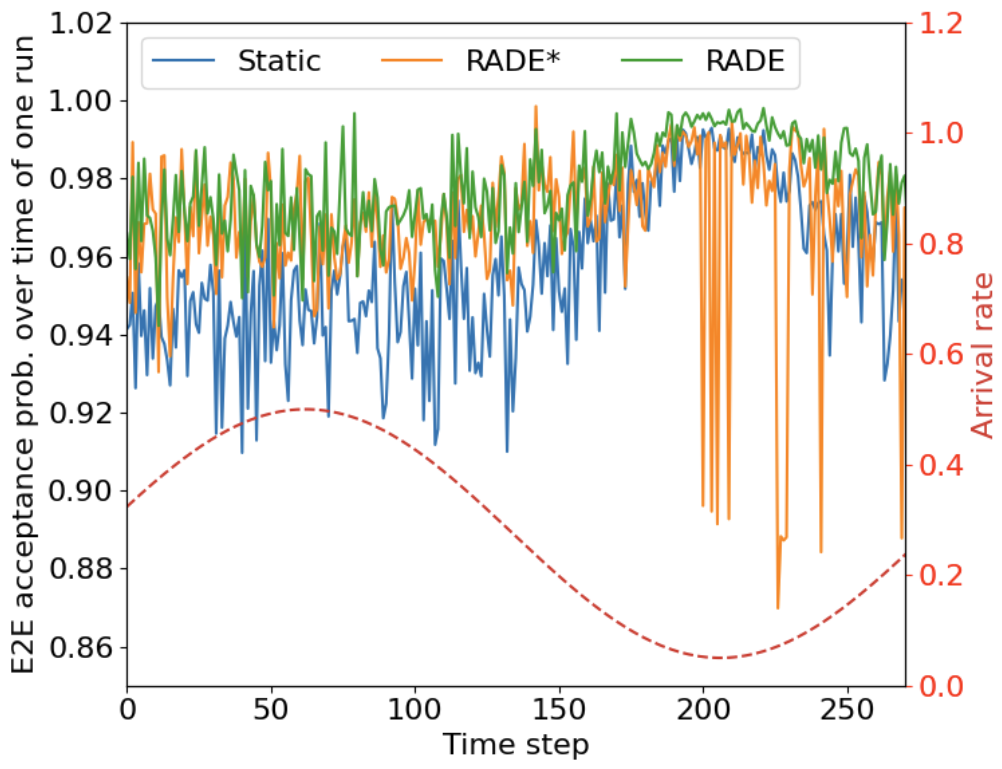}
\caption{One run of E2E acceptance probability over time.}
\label{fig:am_trace}
\end{figure}

\begin{figure}[h]
 \centering
\includegraphics[width=1\columnwidth, trim={0 1.5cm 0 2.8cm}, clip]{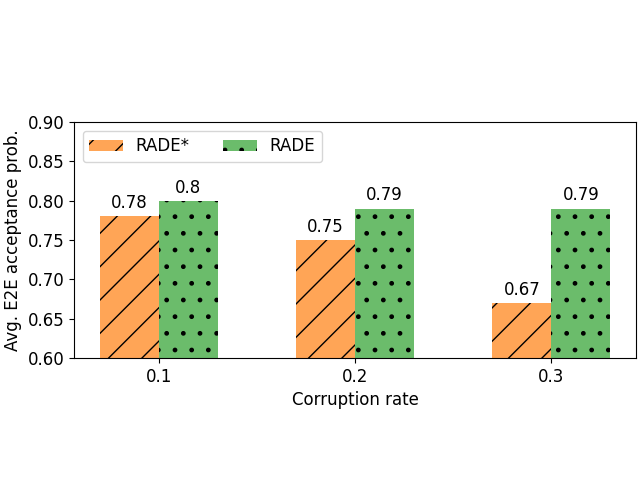}
\caption{Average E2E acceptance probability vs. corruption rate at arrival rate=0.5.}
\label{fig:am_res}
\vspace{-1em}
\end{figure}

Fig.~\ref{fig:am_main} presents the average E2E acceptance probability across different arrival rates for four methods: Static, RADE*, RADE, and OPT. The arrival rates vary between $0.3$, $0.5$, and $0.7$, representing different traffic intensities in the system. Across all arrival rates, the Static method consistently exhibits the lowest average E2E acceptance probability. This suggests that the Static method's one-off trained risk models are less adaptable to varying traffic conditions, leading to sub-optimal performance, particularly under higher traffic loads. RADE* shows improved performance over the Static method, which indicates that dynamically updating the risk models, even without the FIFO memory buffer, leads to a better adaptability than the Static approach. The RADE method, which includes the FIFO memory buffer for maintaining historical feedback, outperforms both Static and RADE*, highlighting the importance of the use of the FIFO buffer in enhancing the robustness and stability of the framework, particularly under low arrival rate conditions. It is interesting to see that the performance of RADE* and RADE becomes nearly identical at higher arrival rates because, under heavy traffic, the system receives a large volume of recent feedback. This plenty of fresh data diminishes the impact of the FIFO memory buffer in RADE, as the most recent observations dominate the learning process. As expected, the OPT method achieves the highest performance, which serves as a theoretical upper bound. In contrast, the Random method performs significantly worse than the other approaches in general, with results of $0.42$, $0.38$, and $0.35$ for arrival rates of $0.3$, $0.5$, and $0.7$, respectively. Due to the consistently low performance of the Random method, we exclude its results from all subsequent figures to focus on the more meaningful comparisons.
Fig.~\ref{fig:am_trace} shows the E2E acceptance probability over time of one run for three methods along with the corresponding arrival rate (the dashed line in red on the secondary y-axis). The Static method exhibits significant fluctuations and lower performance in general. RADE* performs well initially, but its performance degrades sharply when the arrival rate is low (around the $200$-th time step). This drop is likely due to its reliance on only the most recent feedback, making it unstable during periods of low traffic that only sparse feedback are available. RADE consistently outperforms both Static and RADE*, maintaining higher and more stable acceptance probabilities across different arrival rates. The stability and adaptability of RADE are particularly evident during the low arrival rate period, where RADE* struggles.
Fig.~\ref{fig:am_res} illustrates the average E2E acceptance probability for the RADE* and RADE methods under varying corruption rates ($0.1$, $0.2$ and $0.3$). As the rate increases, the performance of RADE* significantly deteriorates, while RADE consistently maintains a higher acceptance probability across all corruption rates. This figure clearly demonstrates again that RADE is more resilient to corrupted feedback compared to RADE*, due to the use of the FIFO memory buffer.

\subsection{Complexity analysis}
The comparison of time complexity is provided in Table~\ref{table:inference_complexity}. Three methods (Static, RADE* and RADE) are analyzed regarding both the complexity of base model training and the decomposition processes, which run in parallel.
Assume the architecture shared by all methods has a training complexity $O(f(D))$ for a single iteration, and $O(g(D))$ for decomposition, where $D$ is the size of the input data, and $f$, $g$ are functions dependent on the architecture.
For the Static method, the time complexity is simply $O(g(D))$, since there is no additional overhead incurred by updates.
For RADE*, the total complexity for $N$-iteration training is $O(N \cdot f(D))$, and the overall complexity is $O(\max(N \cdot f(D), g(D)))$.
Similarly, RADE behaves like RADE* but with the additional use of the FIFO memory buffer. Assume the FIFO memory buffer can hold up to $Q$ data points, the overall complexity is $O(\max(N \cdot f(Q), g(D)))$. Typically, $Q$ is larger than $D$ to ensure effective utilization of the FIFO memory buffer.
In summary, the Static method is the most time-efficient due to its lack of online update. RADE* introduces moderate computational overhead from online update, while RADE is the most computationally intensive, primarily due to the online update performed on the buffered data.

\vspace{-1em}
\begin{table}[h]
\caption{Comparison of time complexity.}
\centering
\begin{tabular}{|c|c|}
\hline
\textbf{Method} & \textbf{Time complexity} \\ 
\hline
\textbf{Static} & $O(g(D))$ \\
\hline
\textbf{RADE*} & $O(\max(N \cdot f(D), g(D)))$ \\
\hline
\textbf{RADE} & $O(\max(N \cdot f(Q), g(D)))$ \\
\hline
\end{tabular}
\label{table:inference_complexity}
\end{table}


\section{conclusion}
\label{sec:conclusion}
In this paper, we introduced RADE, an innovative online learning-decomposition framework designed to address the dynamic nature of SLA management in multi-domain network slicing environments. By continuously updating domain-specific risk models using real-time feedback from domain controllers, RADE effectively adapts to evolving network conditions, overcoming the limitations of traditional static SLA decomposition methods.
Our approach leverages OGD for iterative model updates and FIFO memory buffers to maintain a balance between historical and recent feedback. This combination not only enhances the robustness and stability of the decomposition process but also ensures resilience against challenges such as feedback sparsity and data corruption.
Comprehensive evaluations using an analytic model-based simulator demonstrate that RADE consistently outperforms state-of-the-art static methods in terms of E2E acceptance probability in fluctuating traffic conditions, and robustness under corrupted feedback scenarios. The ablation study further justifies the critical role of each component.
In future work, we will explore the potential of Deep Reinforcement Learning (DRL) that accounts for long-term performance.
Additionally, we aim to investigate scenarios where each domain comprises multiple providers,  introducing selection mechanisms to reflect more realistic and complex multi-domain environments.

\section*{Acknowledgment}
This research was partially funded by the HORIZON SNS JU DESIRE6G project (grant no. 101096466) and the Dutch 6G flagship project ``Future Network Services''.

\bibliographystyle{IEEEtran}
\bibliography{References/references.bib}

\begin{thebibliography}{10}
\providecommand{\url}[1]{#1}
\csname url@samestyle\endcsname
\providecommand{\newblock}{\relax}
\providecommand{\bibinfo}[2]{#2}
\providecommand{\BIBentrySTDinterwordspacing}{\spaceskip=0pt\relax}
\providecommand{\BIBentryALTinterwordstretchfactor}{4}
\providecommand{\BIBentryALTinterwordspacing}{\spaceskip=\fontdimen2\font plus
\BIBentryALTinterwordstretchfactor\fontdimen3\font minus \fontdimen4\font\relax}
\providecommand{\BIBforeignlanguage}[2]{{%
\expandafter\ifx\csname l@#1\endcsname\relax
\typeout{** WARNING: IEEEtran.bst: No hyphenation pattern has been}%
\typeout{** loaded for the language `#1'. Using the pattern for}%
\typeout{** the default language instead.}%
\else
\language=\csname l@#1\endcsname
\fi
#2}}
\providecommand{\BIBdecl}{\relax}
\BIBdecl

\bibitem{ietf-teas-5g-network-slice-application-03}
X.~Geng, L.~M. Contreras, R.~Rokui, J.~Dong, and I.~Bykov, ``{IETF Network Slice Application in 3GPP 5{G} End-to-End Network Slice},'' Internet Engineering Task Force, Internet-Draft draft-ietf-teas-5g-network-slice-application-03, Jun. 2024.

\bibitem{hcltech2023networkslicing}
R.~Swamy and S.~K. M, ``5{G} network slicing,'' HCL Technologies, Tech. Rep., 2023.

\bibitem{iovanna2022networkslicing}
P.~Iovanna, M.~Svensson, A.~Shapin, G.~Bottari, F.~Ubaldi, F.~Ponzini, and M.~Puleri, ``End-to-end network slicing orchestration,'' \emph{Ericsson Technology Review}, vol.~2, 2022.

\bibitem{su2019resource}
R.~Su, D.~Zhang, R.~Venkatesan, Z.~Gong, C.~Li, F.~Ding, F.~Jiang, and Z.~Zhu, ``Resource allocation for network slicing in 5{G} telecommunication networks: A survey of principles and models,'' \emph{IEEE Network}, vol.~33, no.~6, pp. 172--179, 2019.

\bibitem{10011552}
J.~Wang, J.~Liu, J.~Li, and N.~Kato, ``Artificial intelligence-assisted network slicing: Network assurance and service provisioning in 6{G},'' \emph{IEEE Vehicular Technology Magazine}, vol.~18, no.~1, pp. 49--58, 2023.

\bibitem{Vleeschauwer21_SLAdecomposition}
D.~{De Vleeschauwer}, C.~{Papagianni}, and A.~{Walid}, ``Decomposing {SLA}s for network slicing,'' \emph{IEEE Communications Letters}, vol.~25, no.~3, pp. 950--954, March 2021.

\bibitem{SLADNN23}
C.~S.-H. Hsu, D.~D. Vleeschauwer, and C.~Papagianni, ``{SLA} decomposition for network slicing: A deep neural network approach,'' \emph{IEEE Networking Letters}, pp. 1--1, 2023.

\bibitem{8417711}
B.~Khodapanah, A.~Awada, I.~Viering, D.~Oehmann, M.~Simsek, and G.~P. Fettweis, ``Fulfillment of service level agreements via slice-aware radio resource management in 5{G} networks,'' in \emph{2018 IEEE 87th Vehicular Technology Conference (VTC Spring)}, 2018, pp. 1--6.

\bibitem{8931583}
H.~Chergui and C.~Verikoukis, ``Offline {SLA}-constrained deep learning for 5{G} networks reliable and dynamic end-to-end slicing,'' \emph{IEEE Journal on Selected Areas in Communications}, vol.~38, no.~2, pp. 350--360, 2020.

\bibitem{10173672}
A.-C. Maroudis, T.~Theodoropoulos, J.~Violos, A.~Leivadeas, and K.~Tserpes, ``Leveraging graph neural networks for {SLA} violation prediction in cloud computing,'' \emph{IEEE Transactions on Network and Service Management}, vol.~21, no.~1, pp. 605--620, 2024.

\bibitem{9165317}
M.~Iannelli, M.~R. Rahman, N.~Choi, and L.~Wang, ``Applying machine learning to end-to-end slice {SLA} decomposition,'' in \emph{2020 6th IEEE Conference on Network Softwarization (NetSoft)}, 2020, pp. 92--99.

\bibitem{HOI2021249}
S.~C. Hoi, D.~Sahoo, J.~Lu, and P.~Zhao, ``Online learning: A comprehensive survey,'' \emph{Neurocomputing}, vol. 459, pp. 249--289, 2021.

\bibitem{FIFO}
D.~Isele and A.~Cosgun, ``Selective experience replay for lifelong learning,'' \emph{Proceedings of the AAAI Conference on Artificial Intelligence}, vol.~32, no.~1, Apr. 2018.

\end{thebibliography}



%



\end{document}